\documentclass[11 pt, journal, onecolumn]{IEEEtran}
\makeatletter

\usepackage{amsmath,bbold, dsfont,yfonts,amssymb,dsfont,epsfig,psfrag,amsthm,bm,multirow, graphicx,color,mathrsfs}

\usepackage{ mathrsfs,dsfont,yfonts}
\usepackage{bbold}
\usepackage{textcomp}

\def\ps@headings{%
\def\@oddhead{\mbox{}\scriptsize\rightmark \hfil \thepage}%
\def\@evenhead{\scriptsize\thepage \hfil \leftmark\mbox{}}%
\def\@oddfoot{}%
\def\@evenfoot{}}
\makeatother
\usepackage{fullpage}

\usepackage{simplemargins}

\newcommand{\Reals}     {{{\mathrm I\!R}}}  
\newcommand{\Cplx}      {{{\mathsf I}\!\!\!{\mathrm C}}} 

\newcommand{\Cc} {{\mathcal C}}
\newcommand{\Nc} {{\mathcal N}}
\newcommand{\Pulk} {{\underline{{\bm {\mathcal P}}}}}
\newcommand{\Xulk} {{\underline{{\bm {\mathcal X}}}}}
\newcommand{\Sulc} {{\underline{\mathcal S}}}
\newcommand{\sul}  {{\underline s}}
\newcommand{\Aulc} {{\underline{\mathcal A}}}
\newcommand{\aul}  {{\underline a}}
\newcommand{\Hul}  {{\underline H}}
\newcommand{\thetaul}  {{\underline \theta}}
\newcommand{\Zrb}     {{\bf 0}}
\newcommand{\Psib}     {{\bm \Psi}}
\newcommand{\Gammab}     {{\bm \Gamma}}

\newtheorem{proposition}{Proposition}
\newtheorem{theorem}{Theorem}

\newtheorem{lemma}{Lemma}

\begin{document}
\sloppy
\thispagestyle{empty}
\pagestyle{empty}
\title{Exploiting Hybrid Channel  Information for Downlink Multi-User MIMO Scheduling\\}
\author{\IEEEauthorblockN{Wenzhuo Ouyang\IEEEauthorrefmark{1},
Narayan Prasad\IEEEauthorrefmark{2} and
Sampath Rangarajan\IEEEauthorrefmark{2}}\\
\IEEEauthorblockA{\IEEEauthorrefmark{1}Ohio State University, Columbus, OH;}
\IEEEauthorblockA{\IEEEauthorrefmark{2}NEC Labs America, Princeton, NJ}\\
e-mail:  ouyangw@ece.osu.edu, \{prasad, sampath\}@nec-labs.com}
\maketitle

\begin{abstract}
We investigate the  downlink multi-user MIMO (MU-MIMO) scheduling problem in the presence of imperfect Channel State Information at the transmitter (CSIT) that comprises of coarse and current CSIT as well as finer but delayed CSIT. This scheduling problem is  characterized by an intricate `exploitation - exploration tradeoff' between   scheduling the users based on current CSIT for immediate gains, and scheduling them to obtain finer albeit delayed  CSIT and potentially larger future gains. We solve this scheduling problem by formulating a frame based joint scheduling and feedback approach, where in each frame a policy is obtained as the solution to a Markov Decision Process. We prove that our proposed approach can be made arbitrarily close to the optimal and then demonstrate its significant gains over conventional MU-MIMO scheduling.
\end{abstract}

\vspace{1cm}
{\bf Keywords:}
{\small{Channel State Information, Frame-based Policy, Markov Decision Process, Multi-User MIMO}}
\newpage

\section{Introduction}\label{chp:Intro}

Multiple Input Multiple Output (MIMO) technology is essential for the emerging 4G-LTE wireless communication systems. 
In the downlink of such a system, which typically has several active users,
multiple antennas enable simultaneous transmissions to multiple users
by allowing the transmitter (base-station) to transmit (along directions in a signal space) in a manner which ensures that each user can receive its intended signal along at-least one interference-free dimension (a.k.a. the Multi-user MIMO  principle) \cite{MU-MIMO2}. 
The number of active users is generally greater than the maximum supportable number of simultaneous transmissions, which in turn is equal to the number of transmit antennas at the base-station (BS). Consequently, only a subset of users can be selected for the MU-MIMO transmission and hence proper user scheduling is important to achieve a desired network utility (e.g., throughput, fairness).

The usual assumption made in existing literature on MU-MIMO scheduling is that the BS can  obtain the channel state information from all users with sufficient accuracy and with negligible delay. Such information, referred to as the Channel State Information at the Transmitter (CSIT), is crucial to ensure that each scheduled user is not dominated by co-channel interference. 
Typically, the BS obtains CSIT by broadcasting a sequence of pilot symbols, and the users in turn estimate their CSI and feedback their quantized estimates to the BS. This feedback process introduces two sources of imperfections to the CSIT. (1) Estimation and quantization errors (due to limited training and finite codebooks); (2) Delays (due to user processing speeds and less flexible scheduling on the feedback channel).
The impact of erroneous CSIT on MU-MIMO performance has been analyzed in \cite{Caire_fb} and utility maximization for MU-MIMO with erroneous CSIT has been considered in \cite{Caire_fb3}. Delay in the CSIT has hitherto been addressed by using prediction based approaches but their drawback is that they have to assume a model for channel evolution, which is significantly difficult to obtain in practice and they also require the delay to be small enough to allow for useful prediction.

For the scenario where the number of users is small enough so that user scheduling is unnecessary, referred to here as the static scenario,  Maddah-Ali and Tse proposed a scheme, namely the MAT scheme \cite{MAT}, that utilizes CSIT that is error-free albeit completely outdated. Their seminal work revealed that the outdated CSI is an important resource that, when combined with the eavesdropped information at the users, can provide a considerable performance gain in terms of degrees of freedom. Recently, the MAT scheme was extended (for the static scenario) to the hybrid CSIT case by also incorporating coarse and current CSIT \cite{Extended_MAT} to obtain further system gains.
However, in the ubiquitous setting where user scheduling is important, such hybrid CSIT needs to be exploited wisely since it is costly to obtain even delayed but error-free CSI feedback \emph{from all users} for making the scheduling decisions. Indeed, the problem is quite different and more challenging than the static case. User scheduling for the MAT scheme has been considered in \cite{adhikar} but their suggested method is akin to the myopic approach discussed later in this paper.

In this paper, we study MU-MIMO downlink scheduling with hybrid CSIT, erroneous as well as delayed, where the time axis is divided into separate scheduling intervals. We consider the realistic scenario where current and coarse CSIT is obtained from all users while more accurate (not necessarily perfect) but delayed CSIT is obtained \emph{only from the scheduled users}. The scheduling problem is hence characterized by an intricate `exploitation - exploration tradeoff', between   scheduling the users based on current CSIT for immediate gains, and scheduling them to obtain finer albeit delayed  CSIT and potentially larger future gains. The contributions of the paper are listed as follows.


$\bullet$ We tackle the aforementioned `exploitation - exploration tradeoff' by formulating a frame based joint scheduling and feedback approach, where in each frame a policy is obtained as the solution to a Markov Decision Process (MDP), the latter solution being determined via a state-action frequency approach \cite{Eitan}\cite{KrishnaModiano}. 

$\bullet$ We consider a general utility function and associate a virtual queue with each user that guides the achieved utility for that user. Based on MDP solutions and virtual queue evolutions, we show that our proposed frame-based joint scheduling and feedback approach can be made arbitrarily close to the optimal.

In the following we use $(.)^T,(.)^{\dag}$ for the transpose and conjugate transpose, respectively. Moreover, $[\bm A,\bm B]$ and $[\bm A;\bm B]$ are used to denote column-wise and row-wise concatenation of matrices $\bm A$ and $\bm B$, respectively. $\|\bm A\|$ is used to denote the Frobenius norm of the matrix $\bm A$.

\section{System Model and Problem Formulation}


We consider the downlink MU-MIMO scheduling problem with one Base Station (BS) and $N$ users. The BS is equipped with $M_t$ transmit antennas and employs linear transmit precoding. Each user is equipped with a single receive antenna. Time is divided into intervals and we let $\bm h_i[k]\in\;\Cplx^{1\times M_t},\;i=1,\cdots,N$ denote the channel state vector seen by user $i$ in interval $k$. In each interval, a subset of users can be simultaneously scheduled. Further, since each user has only one receive antenna, it  can achieve at-most one degree of freedom (i.e., its average data rate per channel use can scale with SNR as $\log(\rm SNR)$). On the other hand, the system can achieve at-most $M_t$ degrees of freedom in that the total average system rate can scale with SNR as $M_t\log(\rm SNR)$. For notational convenience we assume that in each interval {\em{two users}} can be simultaneously served, hence limiting the achievable system degrees of freedom to 2.
All results can however be extended to the general case without this restriction.
\subsection{Conventional MU-MIMO scheme}\label{sec:convmu}
  Conventional MU-MIMO scheme relies on estimates of the user channel states (that are available at the BS) for the current interval. Indeed, perfect CSIT for the current interval enables the BS to transmit simultaneously to both scheduled users without causing interference at either of them. However, in the absence of perfect CSIT such complete interference suppression via transmitter side processing is no longer possible and  when only very coarse estimates for the current interval are available, conventional MU-MIMO breaks down and in-fact becomes inferior to simple single-user per interval transmission. 

\subsection{Joint Scheduling and Channel Feedback}

\begin{figure}
\centering
\includegraphics[width=2.8in]{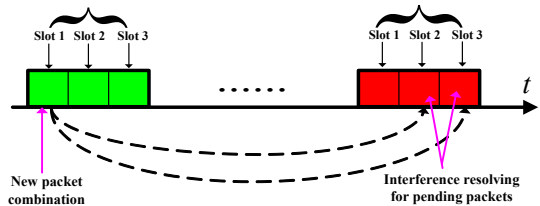}
\caption{Illustration of the scheduling process.}
\label{fig:interval}
\vspace{-10pt}
\end{figure}

We consider a joint scheduling and channel feedback    scheme that builds upon a variant of the extended MAT technique \cite{Extended_MAT}. The extended MAT scheme is recapitulated in Appendix \ref{sec:extEmat}.  Specifically, we assume that coarse quantized channel state estimates from all users for the current interval are available to the BS, along with limited finer albeit outdated quantized channel state estimates. In this context we note that in the FDD downlink only quantized estimates are available to the BS and henceforth unless otherwise mentioned, we will use ``estimates" to mean ``quantized estimates". 
The time duration of interest is divided into intervals with each interval comprising of $3$ slots each. The three slots are mutually orthogonal 
time-bandwidth slices. For convenience, we assume that all three slots in an interval are within the coherence time and coherence bandwidth window so that the channel seen by each user remains constant over the three slots in an interval. At the beginning of the $k^{th}$ interval, whose corresponding slots are denoted by $[k,1],[k,2]$ and $[k,3]$, 
the scheduler broadcasts a short sequence of pilot symbols to all the users.   This sequence enables a coarse  estimation of the wireless channel at each of the $N$ users, which is fed back to the BS after quantization and is denoted by $\hat{\Hul}[k]=\{\hat{{\bm h}}_i[k], i=1, \cdots N\}$, where $\hat{{\bm h}}_i[k]$ denotes the coarse channel estimate obtained from user $i$ for interval $k$. Based on these coarse   estimates, along with its past scheduling and channel state history (formally introduced next),   the scheduler chooses a pair of users to schedule in the current interval, where in the first slot a linear combination of new packets is sent for the selected user pair.
 Data transmission   to the selected user pair in the current interval also contains additional pilots that enable  a finer estimation of the channel states  seen by that user pair over the current interval. Note that such finer estimation is crucial for data detection. 
 However, due to user processing and feedback delays, we assume that (quantized versions of) such finer estimates are not available to the BS during the current interval itself. Because of this constraint,  instead of performing the   transmissions in slots $2$ and $3$ for interference resolution for the packets sent in Slot 1 of the current interval, as would be done in the extended MAT scheme \cite{Extended_MAT}, the BS performs  transmissions for interference resolution for packets sent in Slot 1 of \emph{the prior most recent interval} when the selected user pair  was scheduled.
The scheduling model is illustrated in Fig.~\ref{fig:interval}.

 As mentioned above the scheduler obtains a finer estimate of the channel states  seen by a user pair on the interval in which they are scheduled, at the end of that interval.\footnote{Arbitrary delays in obtaining such finer estimates are also considered later in the paper.}
Let $\thetaul=(u_1,u_2,\kappa)$ represent the $3-$tuple denoting the scheduling decision made for the current interval $k$ such that $u_1,u_2$ denote the selected user pair and $\kappa$ denotes the  index of the prior most recent interval over which that pair was scheduled. We let $\Gamma[k]$ be the collection of the most recently obtained finer channel estimates at the BS for each of the user pairs and their corresponding interval indices, at the start of interval $k$. 
Thus, 
 the set $\Gamma[k]$ takes the form $\Gamma[k]=\big\{(\breve{\bm h}_i[\kappa_{i,j}],\breve{\bm h}_j[\kappa_{i,j}], \kappa_{i,j}), 1 \leq i< j\leq N \big\}$, where $(\breve{\bm h}_i[\kappa_{i,j}],\breve{\bm h}_j[\kappa_{i,j}])$ denote the finer estimates for interval $\kappa_{i,j}$ and $\kappa_{i,j}$ denotes the index of the prior recent-most interval on which pair $i,j$ was scheduled. At the end of that interval (equivalently at the start of interval $k+1$) the set $\Gamma[k+1]$ is obtained by first setting it equal to $\Gamma[k]$ and then updating the $3-$tuple corresponding to the pair $(u_1, u_2)$  selected in interval $k$ to
$(\breve{\bm h}_{u_1}[k], \breve{\bm h}_{u_2}[k],k)$.

The set of user channel states are assumed to be i.i.d. across intervals and the channel states of any two distinct users are assumed to be mutually independent. Given a particular initial rough estimates of the channel states of the user pair selected in interval $k$, $(\hat{{\bm h}}_{u_1}[k], \hat{{\bm h}}_{u_2}[k])$, the distribution of the finer channel estimates  in   the same interval 
is described by the conditional distribution
\begin{IEEEeqnarray}{rCl}
P(\breve{\bm h}_{u_1}[k], \breve{\bm h}_{u_2}[k] \Big| \hat{{\bm h}}_{u_1}[k], \hat{{\bm h}}_{u_2}[k])\label{eq:condprob}
\end{IEEEeqnarray}
where the conditional probability depends on the types of channel estimators, quantization, training times and powers, etc.
We let $\Cc_{\rm coarse}$ ($\Cc_{\rm fine}$) denote the finite sets or codebooks of vectors from which all coarse (fine) estimates are selected. Let $|\Cc_{\rm coarse}|$ and $|\Cc_{\rm fine}|$ denote their respective cardinalities and clearly $|\Cc_{\rm fine}|\geq |\Cc_{\rm coarse}|$.

\subsection{Expected Transmission Rates (Rewards)}\label{sec:exprate}
During the current interval $k$, formed by slots $[k,1],[k,2]\;\&\;[k,3]$, once a pair of users is selected, the scheduler specifies transmit precoding matrices or vectors for each slot in the interval.
\subsubsection{Slot $1$}
 For slot $1$, the overall transmit precoding matrix is denoted by the matrix   $[{\bm W}_{u_1}[k], {\bm W}_{u_2}[k]]$, where
${\bm W}_{u_1}[k], {\bm W}_{u_2}[k]\in\Cplx^{M_t\times 2} $.
Let ${\bm x}_{u_1}[k]={\bm W}_{u_1}[k]{\bm s}_{u_1}[k]$, ${\bm x}_{u_2}[k]={\bm W}_{u_2}[k]{\bm s}_{u_2}[k]$, where ${\bm s}_{u_2}[k],{\bm s}_{u_2}[k]$ denote the $2\times 1$ symbol vectors containing symbols formed using the new packets intended for user $u_1$ and $u_2$, respectively, and where $E[{\bm s}_{u_i}[k]{\bm s}_{u_i}^{\dag}[k]]=\bm I,\;i\in\{1,2\}$. Then, the signal transmitted in slot-1 is ${\bm x}_{u_1}[k]+{\bm x}_{u_2}[k]$ so that the received signals at both users are
\begin{IEEEeqnarray}{rCl}
y_{u_1}[k,1]&={\bm h}_{u_1}[k] ({\bm x}_{u_1}[k]+{\bm x}_{u_2}[k]) + n_{u_1}[k,1],\label{eq:EMAT_Slot11} \\
y_{u_2}[k,1]&={\bm h}_{u_2}[k] ({\bm x}_{u_1}[k]+{\bm x}_{u_2}[k]) + n_{u_2}[k,1].\label{eq:EMAT_Slot12}
\end{IEEEeqnarray}
Note that
 the allocated transmission power for scheduled user $u_i$ is the norm $\|{\bm W}_{u_i}[k] \|^2$. We assume that the maximum average (per-slot) transmission power budget at the BS is $P$. Thus, the corresponding power constraint is
$\|{\bm W}_{u_1}[k] \|^2 + \|{\bm W}_{u_2}[k] \|^2 \leq P$. 
Notice that the precoding matrix $[{\bm W}_{u_1}[k],{\bm W}_{u_2}[k]]$ seeks to facilitate the transmission of new packets to users $u_1$ and $u_2$ and thus must be designed based on the available coarse estimates $(\hat{\bm h}_{u_1}[k], \hat{\bm h}_{u_2}[k])$, since the corresponding finer estimates for that interval are not yet available to the scheduler. Accordingly, we assume that this
precoding matrix  can be obtained as the output of any arbitrary but fixed (time-invariant) mapping from $\Cc_{\rm coarse}\times \Cc_{\rm coarse}$ to  $\;\Cplx^{M_t\times 4}$, when the coarse estimates $(\hat{\bm h}_{u_1}[k], \hat{\bm h}_{u_2}[k])$ are given as an input. Note that assuming the mapping to be fixed is well suited to systems where the so-called ``precoded pilots" are not available so that the choice of precoders needs to be signalled to the scheduled users. A fixed mapping (which is equivalent to one codebook of transmit precoders) then allows for  efficient signaling.


\subsubsection{Slot $2$}

In slot $2$ of the interval, an interference resolving packet for a pending previous transmission involving users $(u_1,u_2)$, sent in interval $\kappa<k$, is transmitted. In particular,
the transmitted signal vector over the $M_t$ antennas is
\begin{IEEEeqnarray}{rCl}
{\bm z}[k,2] \big(\breve{\bm h}_{u_1}[\kappa] \underbrace{ {\bm W}_{u_2}[\kappa]{\bm s}_{u_2}[\kappa]}_{{\bm x}_{u_2}[\kappa]}\big),\nonumber
\end{IEEEeqnarray}
where ${\bm z}[k,2]\in\;\Cplx^{M_t\times 1}$ is a precoding vector. Note that  $\breve{\bm h}_{u_1}[\kappa]{\bm x}_{u_2}[\kappa]$ is a scalar, so the average power constraint
$E[\|{\bm z}[k,2]\breve{\bm h}_{u_1}[\kappa]{\bm x}_{u_2}[\kappa]\|^2]\leq P$ 
can also be written as
$\|{\bm z}[k,2]\|^2 \left\|\breve{\bm h}_{u_1}[\kappa] {\bm W}_{u_2}[\kappa]\right\|^2 \leq P.$  
The received signals in slot $2$ at both users are therefore
\begin{IEEEeqnarray}{rCl}
y_{u_1}[k,2]&={\bm h}_{u_1}[k] {\bm z}[k,2]\big(\breve{\bm h}_{u_1}[\kappa]  {\bm x}_{u_2}[\kappa]\big) + n_{u_1}[k,2]\label{eq:EMAT_Slot21}\IEEEeqnarraynumspace \\
y_{u_2}[k,2]&={\bm h}_{u_2}[k] {\bm z}[k,2] \big(\breve{\bm h}_{u_1}[\kappa]  {\bm x}_{u_2}[\kappa] \big)+ n_{u_2}[k,2].\label{eq:EMAT_Slot22}\IEEEeqnarraynumspace
\end{IEEEeqnarray}

\subsubsection{Slot $3$}

In slot $3$ of the interval, similarly, the transmitted signal is
\begin{IEEEeqnarray}{rCl}
{\bm z}[k,3] \big(\breve{\bm h}_{u_2}[\kappa] \underbrace{ {\bm W}_{u_1}[\kappa]{\bm s}_{u_1}[\kappa]}_{{\bm x}_{u_1}[\kappa]}\big).\nonumber
\end{IEEEeqnarray}
so that the power constraint is
$\|{\bm z}[k,3]\|^2 \left\|\breve{\bm h}_{u_2}[\kappa] {\bm W}_{u_1}[\kappa]\right\|^2 \leq P$.  
The received signals in slot $3$ at both users are therefore
\begin{IEEEeqnarray}{rCl}
y_{u_1}[k,3]&={\bm h}_{u_1}[k]{\bm z}[k,3] \big(\breve{\bm h}_{u_2}[\kappa]  {\bm x}_{u_1}[\kappa]\big) + n_{u_1}[k,3]\label{eq:EMAT_Slot31}\IEEEeqnarraynumspace \\
y_{u_2}[k,3]&={\bm h}_{u_2}[k]{\bm z}[k,3]  \big(\breve{\bm h}_{u_2}[\kappa]  {\bm x}_{u_1}[\kappa]\big) + n_{u_2}[k,3].\label{eq:EMAT_Slot32}\IEEEeqnarraynumspace
\end{IEEEeqnarray}

Notice that the precoding vectors ${\bm z}[k,2],{\bm z}[k,3]$ seek to facilitate the completion of a pending transmission  to users $u_1$ and $u_2$ and thus must be designed based on the available coarse estimates $(\hat{\bm h}_{u_1}[k], \hat{\bm h}_{u_2}[k])$, as well as the available estimates for interval $\kappa$ which are $(\breve{\bm h}_{u_1}[\kappa], \breve{\bm h}_{u_2}[\kappa])$ and $(\hat{\bm h}_{u_1}[\kappa], \hat{\bm h}_{u_2}[\kappa])$. Accordingly, we assume that these two vectors can be obtained as the output of an arbitrary but fixed mapping from $\Cc_{\rm fine}^2\times \Cc_{\rm coarse}^4$ to $\;\Cplx^{M_t\times 2}$. 
 An example of mapping rules to obtain the precoding matrices and vectors  is given later in the section on simulation results.

 Next, in order to compute the average rates (rewards) we assume that the channel state vectors ${{\bm h_{u_i}}}[\kappa],{{\bm h_{u_i}}}[k]$ are known perfectly to user $u_i,i\in\{1,2\}$ (each user of course also knows the quantized estimates it has fed back to the base-station). In addition, user $u_1$ ($u_2$) is also conveyed the finer estimate $\breve{\bm h}_{u_2}[\kappa]$, ($\breve{\bm h}_{u_1}[\kappa]$) via feed-forward signaling before the start of interval $k$. For simplicity, the feedback and feedforward signaling overheads are ignored in this work. 
Then, by the end of slot $3$, from (\ref{eq:EMAT_Slot11}), (\ref{eq:EMAT_Slot21}) and (\ref{eq:EMAT_Slot31}), at user $u_1$, we have
\begin{IEEEeqnarray}{rCl}
y_{u_1}[\kappa,1]-\frac{y_{u_1}[k,2]}{{\bm h}_{u_1}[k] {\bm z}[k,2]}
={\bm h}_{u_1}[\kappa] {\bm x}_{u_1}[\kappa]\quad\quad\quad\quad\quad\nonumber\\
+ ({\bm h}_{u_1}[\kappa]-\breve{\bm h}_{u_1}[\kappa]){\bm x}_{u_2}[\kappa] \quad  \nonumber\\ + n_{u_1}[\kappa,1]-\frac{n_{u_1}[k,2]}{{\bm h}_{u_1}[k] {\bm z}[k,2]},\nonumber\\
y_{u_1}[k,3]=\underbrace{({\bm h}_{u_1}[k] {\bm z}[k,3])}_{\delta_{u_1}[k]} \breve{\bm h}_{u_2}[\kappa] {\bm x}_{u_1}[\kappa] + n_{u_1}[k,3], \label{eq:altobs}\IEEEeqnarraynumspace
\end{IEEEeqnarray}
where the additive noise variables $n_{u_1}[k,1],n_{u_1}[k,2],n_{u_1}[k,3]$ are i.i.d. circularly symmetric complex Gaussian variables with zero-mean and unit variance, $\Cc\Nc(0,1)$.
Notice that the interference term $({\bm h}_{u_1}[\kappa]-\breve{\bm h}_{u_1}[\kappa]){\bm x}_{u_2}[\kappa]$ is independent of the desired signal as well as the additive noise. Letting ${\bm h}^{\rm error}_{u_1}[\kappa]={\bm h}_{u_1}[\kappa]-\breve{\bm h}_{u_1}[\kappa]$, the noise plus interference covariance for user ${u_1}$, denoted by $\Gammab_{u_1}[k]$, is therefore
\begin{IEEEeqnarray}{rCl}
\begin{bmatrix}
1+ \|{\bm h}^{\rm error}_{u_1}[\kappa]{\bm W}_{u_2}[\kappa]\|^2+\frac{1}{|{\bm h}_{u_1}[k] {\bm z}[k,2]|^2} & 0\\
0 & 1
\end{bmatrix}.\nonumber
\end{IEEEeqnarray}
Define {  ${\bm G}_{u_1}[k]=\left[{\bm h}_{u_1}[\kappa] {\bm W}_{u_1}[\kappa];\delta_{u_1}[k] \breve{\bm h}_{u_2}[\kappa] {\bm W}_{u_1}[\kappa]\right]$} and note that ${\bm G}_{u_1}[k]\in\Cplx^{2\times 2}$. Further, let
 $\bm H^{\rm csi}((u_1,u_2),(\kappa,k))=\{\breve{\bm h}_{u_1}[\kappa],\breve{\bm h}_{u_2}[\kappa],\hat{\bm h}_{u_1}[\kappa],\hat{\bm h}_{u_2}[\kappa],\hat{\bm h}_{u_1}[k],\hat{\bm h}_{u_2}[k]\}$ denote the set of channel state information at the scheduler for user pair $u_1,u_2$ over intervals $\kappa,k$.
Then, using (\ref{eq:altobs}) the instantaneous information rate, denoted as $I_{u_1}[k]$ is given by
\begin{IEEEeqnarray}{rCl}
I_{u_1}[k] = \frac{1}{3}\log \left|{\bm I} + \Gammab_{u_1}^{-1}[k]{\bm G}_{u_1}[k]{\bm G}_{u_1}^{\dag}[k]\right|,
 \label{eq:Ru1}
\end{IEEEeqnarray}
where the fraction $1/3$ is to account for the fact that three slots are needed to obtain this rate.
Then,  (an optimistic value for) the average information rate that can be achieved via rateless coding (cf. \cite{MUMIMO_sch3}) is given by
\begin{IEEEeqnarray}{rCl}
R_{u_1}^{\rm opt}[k] = E\left[I_{u_1}[k]\;\middle\vert\;\bm H^{\rm csi}((u_1,u_2),(\kappa,k))\right].
 \label{eq:Ru1ea}
\end{IEEEeqnarray}
A more conservative rate that is appropriate for conventional coding, denoted as $R_{u_1}^{\rm conv}[k]$,  is given by
{ {\begin{IEEEeqnarray}{rCl}
 r_{\thetaul,u_1}\left(1- \Pr\left(I_{u_1}[k]<r_{\thetaul,u_1}\;\middle\vert\;\bm H^{\rm csi}((u_1,u_2),(\kappa,k))\right)\right),\label{eq:Ru1eb}\IEEEeqnarraynumspace
\end{IEEEeqnarray}}}
where $r_{\thetaul,u_1}$ denotes the rate assigned (using any fixed mapping) to user $u_1$ in  $\thetaul$ before transmission of new packets for the pair $(u_1,u_2)$ in interval $\kappa$, based on the available coarse estimates $\hat{\bm h}_{u_1}[\kappa],\hat{\bm h}_{u_2}[\kappa]$.
The rates corresponding to (\ref{eq:Ru1ea}) or (\ref{eq:Ru1eb}) can be derived in a similar manner for user $u_2$.

Note that in deriving the average rate in (\ref{eq:Ru1ea}) or (\ref{eq:Ru1eb}) we have assumed a simple albeit sub-optimal filtering at the user to suppress the interference from the transmission intended for the co-scheduled user. For completeness, we provide the average rate expressions for the case when the user employs the optimal linear filter and  for brevity we only consider the optimistic rate for user $u_1$. Towards this end, we collect the observations received by user $u_1$ as

{{ {\begin{IEEEeqnarray}{rCl}
  \begin{bmatrix}
\nonumber y_{u_1}[\kappa,1]\\y_{u_1}[k,2]\\y_{u_1}[k,3]
\end{bmatrix} =  {\bm F}_{u_1}[k] {\bm x}_{u_1}[\kappa] +
 \tilde{\bm  F}_{u_1}[k]  {\bm x}_{u_2}[\kappa] +
 \begin{bmatrix}
 n_{u_1}[\kappa,1]\\n_{u_1}[k,2]\\n_{u_1}[k,3]
\end{bmatrix},\label{eq:rxob}\IEEEeqnarraynumspace
\end{IEEEeqnarray}}}}
where
{{ {\begin{IEEEeqnarray}{rCl}
\nonumber {\bm F}_{u_1}[k]= \begin{bmatrix}
 {\bm h}_{u_1}[\kappa] \\ {\bm 0} \\ \delta_{u_1}[k]\breve{\bm h}_{u_2}[\kappa]
\end{bmatrix}, \tilde{\bm  F}_{u_1}[k]= \begin{bmatrix}
 {\bm h}_{u_1}[\kappa] \\  {\bm h}_{u_1}[k]{\bm z}[k,2]\breve{\bm h}_{u_1}[\kappa]\\{\bm 0}
\end{bmatrix}
\end{IEEEeqnarray}}}

For this model, we can determine the  instantaneous information rate that can be achieved via optimal filtering using (\ref{eq:Ru1}) but where
 where $\Gammab_{u_1}[k]=  {\bm I} + \tilde{\bm  F}_{u_1}[k]{\bm W}_{u_2}[\kappa]{\bm W}_{u_2}^{\dag}[\kappa]\tilde {\bm F}_{u_1}^{\dag}[k]$ and
  ${\bm G}_{u_1}[k]= {\bm F}_{u_1}[k]{\bm W}_{u_1}[\kappa]$.
The average information rate   can then determined as before using
 (\ref{eq:Ru1ea}).

%

We assume that either conventional coding is employed for all users or rateless coding is employed and accordingly let $R_{u_i}[k],\;1\leq i\leq 2$  denote the average rate, henceforth referred to also as the service rate,  obtained over interval $k$. We also note here that the scheduling scheme (policy) is preceded by an initial set-up phase comprising of $N(N-1)/2$ intervals in which new packets are transmitted successively to each user pair without any accompanying interference resolution packets. For notational convenience, we assume that the scheduling policy starts operating from interval with index $0$ using the initial set $\Gamma[0]$ determined by the set-up phase. 

\subsection{Incorporating one-shot transmissions and feedback delays}
We first consider the case of one-shot transmissions.
To enable one-shot transmission of packets to any pair in any interval $k$, we define an action $\thetaul$ in which $u_1,u_2$ is the pair but
 $\kappa=\phi$ to capture the fact that the intended transmission is one-shot and hence does  not seek to resolve any pending previous transmission.
 Then, in all three slots of that interval transmission is done as in conventional MU-MIMO relying only on the available current estimates
  $\hat{\Hul}[k]$. In particular, a transmit precoder $[\bm w_{u_1}[k],\bm w_{u_2}[k]]\in\;\Cplx^{M_t\times 2}$ is formed based on $\{\hat{\bm h}_{u_1}[k],\hat{\bm h}_{u_2}[k]\} $ using a technique such as zero-forcing \cite{MU-MIMO1}.  Defining
  $I_{u_1}^{\rm one-shot}[k]= \log\left(1+ |\bm h_{u_1}[k]\bm w_{u_1}[k]|^2/(1+|\bm h_{u_1}[k]\bm w_{u_2}[k]|^2)\right)$,
  the corresponding average rates obtained for user $u_1$ (similarly for user $u_2$) are given by
   \begin{IEEEeqnarray}{rCl}
   E\left[I_{u_1}^{\rm one-shot}[k]\;\middle\vert\;\hat{\bm h}_{u_1}[k],\hat{\bm h}_{u_2}[k]\right],
 \label{eq:Ru1en}
\end{IEEEeqnarray}
or
\begin{IEEEeqnarray}{rCl}
 r_{\thetaul,u_1}\left(1- \Pr\left(I_{u_1}^{\rm one-shot}[k]<r_{\thetaul,u_1}\;\middle\vert\;\hat{\bm h}_{u_1}[k],\hat{\bm h}_{u_2}[k]\right)\right).
 \nonumber 
\end{IEEEeqnarray}
    In addition at the end of interval $k$, we simply set $\Gamma[k+1]=\Gamma[k]$ since no pending packets are completed or introduced. 

    Recall that so far we have assumed that upon choosing action $\thetaul$ for interval $k$, the finer estimates
     $\breve{\bm h}_{u_1}[k],\breve{\bm h}_{u_2}[k]$ are available at the start of interval $k+1$ (representing a unit delay).
     In practical systems there can be a delay of several intervals in obtaining such finer estimates. Assuming that these delays are fixed and known in advance, they can be accommodated by expanding the definition of a state. In particular, we can define $4-$tuples such as
      $(i,j,\kappa_{i,j},d_{i,j})$ where $d_{i,j}\geq 0$ measures the remaining delay after which finer estimates  $\breve{\bm h}_{i}[\kappa_{i,j}],\breve{\bm h}_{j}[\kappa_{i,j}]$ will be available. At any interval $k$ selecting the
action $(i,j,\kappa_{i,j},d_{i,j})$ with $d_{i,j}>0$ ($d_{i,j}=0$) constrains the interference resolution to be based only on the coarse estimates  $\hat{\bm h}_{i}[\kappa_{i,j}],\hat{\bm h}_{j}[\kappa_{i,j}],\hat{\bm h}_{i}[k],\hat{\bm h}_{j}[k]$ (on both coarse and fine estimates $\bm H^{\rm csi}((i,j),(\kappa_{i,j},k))$). Upon selecting this action the $4-$tuple in
$\Gamma[k+1]$ corresponding to the pair $i,j$ is set to be $(i,j,k,d_{i,j}=D_{i,j})$ where  $D_{i,j}$ is the maximum  delay (starting from $k+1$) after which the finer estimates will be available. If that action is not selected, it is updated in $\Gamma[k+1]$ as $(i,j,\kappa_{i,j},d_{i,j}=\max\{0,d_{i,j}-1\})$.
For convenience in exposition the aforementioned two extensions are not considered below.

\subsection{System State and Throughput Region}
Define the system state at the start of an interval $j$ as $S[j]=\{ \Gamma[j], \hat{\Hul}[j] \}$ and let $\thetaul[j]$ denote the decision (action) taken in that interval. Then,  at each interval $k$,
  a scheduling policy $\psi$   takes as input all the history up-to interval $k$, comprising of  states $\{S[j]\}_{j=0}^k$ and all decisions $\{\thetaul[j]\}_{j=0}^{k-1}$, to output a decision $\thetaul[k]$.
Under a particular policy $\psi$, the throughput of the $n^{th}$ user is denoted as
\begin{IEEEeqnarray}{rCl}\label{eq:thrregion}
r_n^{\psi}=\lim_{J \rightarrow \infty} \frac{1}{J} \sum_{t=0}^{J-1} E\big[R_n^{\psi}[t] \big]\;\forall\;n,
\end{IEEEeqnarray}
where $R_n^{\psi}[t]=R_n[t]\bm 1 (n \in \thetaul[t])$ and
the expectation is over the initial state  and the evolution of the states and decisions in the subsequent intervals. Note that in (\ref{eq:thrregion}) for simplicity we have assumed that the limit exists for the selected policy. In case the limit does not exist, we can consider any sub-sequence for which the limit exists. Let $\Psib$ be the set of all  policies.
The throughput region that is of interest to us is defined as the closure of the convex hull of the throughput vectors achievable under all policies in $\Psib$, i.e.,
\begin{IEEEeqnarray}{rCl}
\Lambda= CH \big\{{\bm r}: \exists \psi \in\Psib \ s.t., {\bm r}={\bm r}^{\psi} \big\}, \nonumber
\end{IEEEeqnarray}
where $CH\{\cdot\}$ denotes closure of the convex hull. For each throughput vector ${\bm r}$, we obtain a utility value $U({\bm r})$, where $U(\cdot)$ is the non-negative component-wise non-decreasing and concave utility function. For convenience, we also assume that the utility is continuous (and hence uniformly continuous) in the closed hypercube $[0,b]^N$ for each finite $b\in\Reals_+$.
The objective then is to   maximize the network utility within the throughput region, i.e., $ \max_{\bm r:\bm r\in\Lambda} U(\bm r)$.

\section{Optimal frame-based scheduling policy}

In this section, we propose a frame based policy that achieves a utility   arbitrarily close to the optimal. In this policy, the time intervals are further grouped into separate frames, where each frame consists of $T$ consecutive intervals. The scheduling decisions in each frame are based on a set of virtual queues that guide the achieved system utility towards optimal, as specified next.

\subsection{Virtual Queue and Virtual Arrival Process}

To control the achieved utilities of different users, a virtual queue is maintained for each user, denoted as $Q_n[k], k=0,1,\cdots\;\&\;n=1, \cdots, N$.
At the beginning of the $\tau^{th}$  frame comprising of intervals $\{\tau T,\cdots,(\tau+1) T-1\}$, where $\tau \in\{0,1,2,\cdots\}$, the following optimization problem is solved at the scheduler
\begin{IEEEeqnarray}{rCl}
\max_{\bm r:\Zrb\preceq {\bm r}\preceq r_{\rm max}\bm {1}} V\cdot U({\bm r})-\sum_{n=1}^N Q_n[\tau T]  r_n, \label{eq:Aopt}
\end{IEEEeqnarray}
where $r_{\rm max},V$ are positive constants that can be freely chosen and whose role will be revealed later.
We let ${\bm r}^*[\tau]$ be the optimal solution to the above problem. Then, the virtual arrival rate for user $n$ is set as  $r^*_n[\tau]$ in each interval  in the $\tau^{th}$ frame.
A scheduling policy, $\psi^*_{\bm Q[\tau T]}$, is determined and implemented based on the virtual queue length $\bm Q[\tau T]$ obtained at the beginning of that frame. Letting $R_{n}^{\Psi^*_{{\bm Q[\tau T]}}}[k]$ denote the service rate of user $n$ in each interval $k$ in the $\tau^{th}$ frame under this policy, the virtual queue is then updated as
\begin{IEEEeqnarray}{rCl}
Q_n[k+1]=\left(Q_n[k]- R^{\Psi^*_{\bm Q[\tau T]}}_n[k] \right)^+ + r^*_n[\tau],
\end{IEEEeqnarray}
for all $\tau T\leq k\leq (\tau+1) T-1$ and each user $n$ and where $(x)^+=\max\{0,x\}$ with $Q_n[0]=0$ for all $n$.

%

\subsection{State-action frequency approach}

We now determine the policy $\Psi^*_{\bm Q[\tau T]}$ employed in the $\tau^{th}$ frame.
Notice that while the definition of the system state adopted thus far allows us to compactly describe any policy, one associated drawback is that the number  of states becomes countably infinite. 
Fortunately, there is one aspect that we can exploit. Note that the average rates obtained upon scheduling a pair of users $i,j$ on any interval $k$ depends only on the corresponding coarse and fine channel estimates in interval $\kappa_{i,j}$ (which we recall denotes the prior recent-most interval over which that pair was scheduled) and the coarse channel estimates in interval $k$ but not on those interval indices. Then, to analyze the average rates offered by any policy, it suffices to define a finite set of states, $\Sulc$, as follows. A state $\sul\in\Sulc$ is defined as a particular choice ${\bm h}_{i}^{\rm p,fine},{\bm h}_{j}^{\rm p,fine},{\bm h}_{i}^{\rm p,coarse},{\bm h}_{j}^{\rm p,coarse},{\bm h}_{i}^{\rm c,coarse},{\bm h}_{j}^{\rm c,coarse}$ of  coarse and fine channel estimates  for each pair $i,j$, where the superscripts $p,c$ denote past and current estimates, respectively. Consequently there are $|\Sulc|=\left(|\Cc_{\rm fine}|^2|\Cc_{\rm coarse}|^2\right)^{\frac{N(N-1)}{2}}|\Cc_{\rm coarse}|^N$ number of states.
Note that a state $S[k]$ in the previous definition would map to state $\sul\in\Sulc$ which has the choice $\breve{\bm h}_{i}[\kappa_{i,j}],\breve{\bm h}_{j}[\kappa_{i,j}],\hat{\bm h}_{i}[\kappa_{i,j}],\hat{\bm h}_{j}[\kappa_{i,j}],\hat{\bm h}_{i}[k],\hat{\bm h}_{j}[k]$ for each pair $i,j$.
  A finite set of actions, $\Aulc$, is defined next to be the collection of all possible user pairs so that any $\aul\in\Aulc$ uniquely identifies a user pair. Let $ P(\sul \big| \sul', \aul)$ denote the transition probability, which we note can be determined using (\ref{eq:condprob}) and the facts that the finer past estimates of pairs not in $\aul$ do not change and the current coarse estimates are i.i.d. across intervals.
    Letting $\Pulk(\Aulc)$ define the set of all probability distributions on $\Aulc$, any policy can be defined as a mapping which at each interval $k$ takes as input all the history up-to interval $k$, comprising of  states $\{\sul[j]\}_{j=0}^k$ and all actions $\{\aul[j]\}_{j=0}^{k-1}$, to output  a distribution in $\Pulk[\Aulc]$ from which the action $\aul[k]$ can be generated.
  A stationary policy is one which at any  interval $k$ considers only the state $\sul[k]$ to output a distribution in $\Pulk[\Aulc]$ and where the output distribution depends only on the state  $\sul[k]$ but not on the interval index $k$.  Under any stationary policy the sequence $\{\sul[k]\}_{k=0}^{\infty}$ is a {\em Markov Chain}.

With these definitions in hand, 
 we let $R_n(\sul,\aul)$ denote the achieved transmission rate for user $n$  when action $\aul$ is taken and the system state is $\sul$. 
Denote the state action frequencies by $\{x(\sul,\aul)\}_{\sul\in\Sulc,\aul\in\Aulc}$, where we note that each $x(\sul,\aul)$ lies in the unit interval $[0,1]$ and   represents the frequency that the system state is at $\sul$ and action $\aul$ is taken. The state action frequencies need  to satisfy the normalization equation
\begin{IEEEeqnarray}{rCl}
\sum_{\sul,\aul} x(\sul,\aul)=1, \nonumber
\end{IEEEeqnarray}
and the balance equation
\begin{IEEEeqnarray}{rCl}
\sum_{\aul} x(\sul,\aul)=\sum_{\sul', \aul} P(\sul \big| \sul', \aul) x(\sul', \aul). \nonumber
\end{IEEEeqnarray}
The above two equations form a state-action polytope $\Xulk$ and let ${\bm x}$ denote any vector of state action frequencies lying in $\Xulk$.
We next define a rate region as
{ {\begin{IEEEeqnarray}{rCl}\label{eq:Newregion}
\tilde{\Lambda}=\{{\bm R}: R_n=\sum_{\sul}\sum_{\aul}R_n(\sul,\aul) x(\sul,\aul),\;\forall\;n \;\&\;{\bm x}\in \Xulk\}.\IEEEeqnarraynumspace
\end{IEEEeqnarray}}}

Then, given the virtual queue length ${\bm q}={\bm Q}[\tau T]$ we
consider the following linear program (LP),
\begin{IEEEeqnarray}{rCl}
\max_{\bm x} &\hspace{5pt} \sum_{\sul,\aul} {\bm q}^T{\bm R}(\sul,\aul)  x(\sul,\aul) \nonumber\\
s.t. & \hspace{8pt} {\bm x} \in \Xulk. \label{eq:LP}
\end{IEEEeqnarray}


We use ${\bm x}^*$ to denote an optimal solution to the linear program and define $\bm R^*=[R_1^*,\cdots,R_N^*]^T$, where
\begin{IEEEeqnarray}{rCl}
R_n^*=\sum_{\sul}\sum_{\aul}R_n(\sul,\aul) x^*(\sul,\aul),\;\forall\;n.\label{eq:Rnstar}
\end{IEEEeqnarray}
 Using the Bayesian rule, we can identify the corresponding stationary policy $\Psi^*_{{\bm Q}[\tau T]}$, which at any interval $k$ in the $\tau^{th}$ frame first maps the state $S[k]$ to its counterpart $\sul\in\Sulc$. Then, if $\sum_{\aul'} x^*(\sul, \aul')>0$, it chooses action $\aul$ using the  probabilistic rule
\begin{IEEEeqnarray}{rCl}
P(\text{pick $\aul$ at state $\sul$})=\frac{x^*(\sul,\aul)}{\sum_{\aul'} x^*(\sul, \aul')},\;\forall\;\aul\in\Aulc. \nonumber
\end{IEEEeqnarray}
On the other hand, if $\sum_{\aul'} x^*(\sul, \aul')=0$, it chooses action $\aul$ arbitrarily. 
Let ${\bm R}^{\rm frame}[k],\tau T\leq k\leq (\tau+1)T-1$, denote the service rate vectors obtained under this policy for the intervals in the $\tau^{th}$ frame.

We list the following results which can be obtained using those that have been derived before for weakly communicating Markov Decision Processes \cite{Eitan},\cite{KrishnaModiano}.
\begin{lemma}\label{prop:region}
 The region  $\Lambda$ defined in (\ref{eq:thrregion}) is identical to the region $\tilde{\Lambda}$ defined in (\ref{eq:Newregion}). Further, for each frame $\tau$ and any given $ {\bm Q[\tau T]}$, an optimal solution to the LP in (\ref{eq:LP}) can be found for which the corresponding  policy $\Psi^*_{{\bm Q}[\tau T]}$ is also deterministic.
  \end{lemma}
  Henceforth, we assume  $\Psi^*_{{\bm Q}[\tau T]}$ to be also deterministic.
\begin{lemma}\label{lem:mixing}
 For arbitrarily fixed $\delta>0$ there exists a large enough   frame length $T_o$ and constants $\gamma,\beta$ such that for each frame length $T\geq T_o$ and all $\bm Q[\tau T]$
 { {\begin{IEEEeqnarray}{rCl}
{\rm Pr}\left(\left\|\frac{1}{T}\left(\sum_{j=0}^{T-1}\bm R^{\rm frame}[\tau T+j]\right)- \bm R^*\right\| >\delta \middle\vert\bm Q[\tau T]\right)\nonumber\\\leq \gamma\exp(-\beta T).\IEEEeqnarraynumspace
\end{IEEEeqnarray}}}
   \end{lemma}

%
%
%
%
%

\subsection{Optimality of the frame-based policy}

Define Lyapunov function $L(\bm Q[\tau T])=\frac{1}{2} \sum_{n=1}^N Q^2_n[\tau T]$. Then the $T$-step average Lyapunov drift is expressed as
{ {\begin{IEEEeqnarray}{rCl}
\Delta_T({\bm Q}[\tau T])=\frac{1}{T} E\left[L(Q[(\tau+1)T])-L(Q[\tau T])\mid {\bm Q}[\tau T] \right],\nonumber
\end{IEEEeqnarray}}}
where the expectation is over the initial states at interval $\tau T$ induced by the policies adopted in the previous frames and the evolution of the states and decisions in the $\tau^{th}$ frame
 under the  policy $\Psi^*_{{\bm Q}[\tau T]}$.
 Our first result is the following. 
 \begin{proposition}\label{prop:P1}
 For any given $\epsilon>0$, there exists a frame length $T_o$ such that for all frame lengths $T\geq T_o$
  the $T$-step average Lyapunov drift can be bounded as
    { {\begin{IEEEeqnarray}{rCl}
\Delta_T({\bm Q}[\tau T])\leq BT  - \sum_{n=1}^N Q_n[\tau T]R_n +  \sum_{n=1}^N Q_n[\tau T]r^*_n[\tau],\IEEEeqnarraynumspace\label{eq:Tub}
\end{IEEEeqnarray}}}
 where $B$ is a constant and ${\bm R}=[R_1,\cdots,R_N]^T$ is any vector such that ${\bm R}+\epsilon{\bf 1}\in \Lambda$.
  \end{proposition}
\proof Proved in Appendix \ref{app:prop1}.\endproof

Consider the $\epsilon$-interior of $\Lambda$, i.e., $\Lambda_{\epsilon}=\{\bm R:  \bm R+\epsilon \bm 1 \in \Lambda\}$. Denote ${\bm r}^{\rm opt}_{\epsilon}$ as the optimal value of the following optimization problem.

\begin{IEEEeqnarray}{rCl}
&\max \hspace{3pt} U({\bm r}) \nonumber \\
&\hspace{5pt}\text{s.t. \  } {\bm r}\in \Lambda_{\epsilon};\bm r\preceq r_{\rm max}\bm 1. \nonumber
\end{IEEEeqnarray}
Our main result is the following. 
  \begin{theorem}\label{prop:P2}
 For any given $\epsilon>0$, there exists a $T_o$ such that for all frame lengths $T\geq T_o$
    \begin{IEEEeqnarray}{rCl}
 \liminf_{J \rightarrow \infty}U\left( \frac{1}{J} \sum_{t=0}^{J-1} E\left[{\bm R}^{\rm frame}[t]\right] \right)\geq U({\bm r}^{\rm opt}_{\epsilon}) - BT/V.\nonumber
\end{IEEEeqnarray}
  \end{theorem}
\proof Proof Sketch in Appendix \ref{app:thm1}.\endproof
Thus, by choosing $\epsilon$, framelength $T$ and parameters $V,r_{\rm max}$ appropriately, our frame based policy can be made arbitrarily close to optimal.

For comparison we will use the conventional MU-MIMO scheduling described in Section \ref{sec:convmu}. In addition, we also use the following {\em myopic} policy.
This policy operates in a manner similar to the frame based policy but with the following important differences.
Firstly, the frame-length is set as $T=1$ so that the arrival rates are computed at the start of each interval and the virtual queues are updated at the end of that interval.
Then, at each interval $k$ the current state $S[k]$ is mapped to its image $\sul\in\Sulc$. Considering the queue length $\bm q=\bm Q[k]$, the action $\hat{\aul}=\arg\max_{\aul\in\Aulc}\bm q^T\bm R(\sul,\aul)$ is selected.
Clearly, this policy does not consider the transition probabilities (and the possible future evolutions) at all while deciding an action. Nevertheless, as seen in the following section, this policy indeed offers a competitive performance.
\section{Simulation Results}
  We consider a narrowband downlink with four single-antenna users that are served by a BS equipped with four transmit antennas. All users are assumed to experience an identical (large scale fading) pathloss factor $\delta$ and thus see an identical average SNR, which models the physical scenario in which all users are equidistant from the BS.  Further, we model the small-scale fading seen by each user as Rayleigh fading so the channel response vector of each user is assumed to have i.i.d. $\Cc\Nc(0,\delta^2)$ elements. Consequently the normalized channel response vector (i.e., channel direction) is isotropically distributed in $\;\Cplx^{4\times 1}$. 
  Moreover, the channel response vectors  evolve  independently across intervals and are independent across users.
    In the following simulations, each user quantizes its channel norm and channel direction separately. In particular, the channel norm is quantized using a scalar quantizer which for simplicity we assume to be identical for both fine and coarse estimates. On the other hand, to quantize the channel direction, in order to obtain the finer estimate, the quantization codebook  used  comprises of a set independently generated instances of isotropic vectors in $\;\Cplx^{4\times 1}$ (a.k.a. random vector codebook), where we note that for large codebook sizes  random vector codebooks have been shown to be a good choice for both SU-MIMO and conventional MU-MIMO. The quantization of the channel direction to obtain the coarser estimate is accomplished using Grasmannian codebooks.

   Before offering our results, we consider an interval $k$ and decision $\thetaul$ and describe the mapping rules alluded to in Section \ref{sec:exprate}.
   We determine a good direction (i.e., unit-norm beamforming vector) for multicasting using the alternating optimization based multicast beamforming design algorithm \cite{zhuP} that takes only the coarse estimates $\hat{\bm h}_{u_1}[k]$ and $\hat{\bm h}_{u_2}[k]$ as inputs and set  $\frac{\bm z[k,2]}{\|\bm z[k,2] \|}$ and $\frac{\bm z[k,3]}{\|\bm z[k,3] \|} $ to be equal to this direction.   The precoding matrix $\bm W_{u_1}[\kappa]$ is   obtained by extending the naive zero-forcing design of conventional MU-MIMO   to the model in (\ref{eq:altobs}). In particular at interval $\kappa$ the BS naively assumes that coarse estimates $\hat{\bm h}_{u_1}[\kappa],\hat{\bm h}_{u_2}[\kappa]$ it has are indeed equal to their respective exact channels (and hence their respective finer estimates). Then, at any future interval $k$ (the knowledge of $k$ is not assumed during interval $\kappa$) when pair $(u_1,u_2)$ is next scheduled, under the naive assumption (\ref{eq:altobs}) would reduce to
   \begin{IEEEeqnarray}{rCl}
y_{u_1}[\kappa,1]-\frac{y_{u_1}[k,2]}{{\bm h}_{u_1}[k] {\bm z}[k,2]}
=\hat{\bm h}_{u_1}[\kappa] {\bm x}_{u_1}[\kappa]+  n_{u_1}[\kappa,1]\nonumber\\-\frac{n_{u_1}[k,2]}{({\bm h}_{u_1}[k] {\bm z}[k,2])},\nonumber\\
\frac{y_{u_1}[k,3]}{({\bm h}_{u_1}[k] {\bm z}[k,3])}= \hat{\bm h}_{u_2}[\kappa] {\bm x}_{u_1}[\kappa] + \frac{n_{u_1}[k,3]}{({\bm h}_{u_1}[k] {\bm z}[k,3])}. \label{eq:altobs2}\IEEEeqnarraynumspace
\end{IEEEeqnarray}
   To remove dependence on $k$, all noise covariances are averaged so that (\ref{eq:altobs2}) reduces to a point-to-point MIMO channel with channel matrix $[\hat{\bm h}_{u_1}[\kappa] ; \hat{\bm h}_{u_2}[\kappa]]$ and noise covariance ${\rm diag}\{1+E[1/|{\bm h}_{u_1}[k] {\bm z}[k,2]|^2],E[1/|{\bm h}_{u_1}[k] {\bm z}[k,3]|^2\}$. Notice however that due to the power constraints  these expected values in turn depend on the choice of precoders $\bm W_{u_1}[\kappa], \bm W_{u_2}[\kappa]$. As a further simplification, we fix these expected values to be suitable scalars which are determined   offline.
   The precoder $\bm W_{u_1}[\kappa]$ can now be obtained using the standard point-to-point MIMO precoder design algorithm \cite{Tse}.
   The precoder $\bm W_{u_2}[\kappa]$ is computed in an analogous manner. Finally, the norms of the precoding vectors
    are fixed as $ \|\bm z[k,2] \| = \frac{\sqrt{P}}{\|\breve{\bm h}_{u_1}[\kappa]\bm W_{u_2}[\kappa]\|}$ and $ \|\bm z[k,3] \| = \frac{\sqrt{P}}{\|\breve{\bm h}_{u_2}[\kappa]\bm W_{u_1}[\kappa]\|}$.

In Fig. \ref{fig:simul}  we compare the sum rate utility obtained using conventional MU-MIMO that only uses the current CSI with that obtained using the myopic scheduling that uses only the delayed CSI (EMAT with delayed) and  the myopic scheduling that uses the hybrid CSI (EMAT with hybrid), where   for the latter two schemes the average rates are computed assuming both the sub-optimal and the optimal filtering. In all cases the channel norms were assumed to be perfectly quantized whereas a 2-bit coarse codebook and 5-bit fine codebook were employed to quantize the channel directions, respectively.
As seen from the figure, the conventional MU-MIMO gets interference limited and the policy using the finer albeit delayed CSI offers significant gains, which are further improved by utilizing the hybrid CSI. The improvement is more marked upon using optimal filtering. 

In Fig. \ref{fig:simul2}  we consider the same setup as in the previous figure but now compare the sum rate utility obtained using   the myopic scheduling that uses  the hybrid CSI  along with the optimal filtering, for different codebook sizes. In particular, in all cases the channel norms were assumed to be perfectly quantized and a 2-bit coarse codebook was employed. Four different codebook sizes (5, 10, 12, and 16 bits) for the fine codebook were employed and compared along with the case when perfect delayed CSI is available to the BS.
As seen from the figure, to capture the promised multiplexing gains the codebook sizes must scale sufficiently fast with SNR. We note here that the MAT and EMAT schemes   have been designed with the goal of achieving degree of freedom improvements, where aligning (confining) interference to a low dimensional subspace is the paramount concern. The substantial gap compared to the perfect delayed CSI performance observed at a fixed (finite) SNR can be alleviated via proper precoder design that is optimized for a finite SNR. We emphasize that the precoder optimization we undertook to produce these set of results were limited and adhered fully to the EMAT framework.

We also compared the sum rates obtained using our proposed policy and the myopic one, respectively, for a simpler examples having   fewer number of states. We found that for well designed quantization codebooks, the myopic policy performs very close to the optimal frame based policy. This observation coupled with the fact that the complexity of the myopic policy scales much more benignly with the system size, makes it well suited to practical implementation. 

\section{Conclusions}
We considered the DL MU-MIMO scheduling problem with hybrid CSIT and  proposed an optimal frame-based joint scheduling and feedback approach. 
There are two  important and interesting issues that are the focus of our current research. The foremost one pertains to the exceedingly large number of states that are needed to accommodate practical system sizes which makes implementation of the frame based policy challenging even upon using commercial LP solvers. While the sparse nature of these linear programs can indeed be exploited, an efficient and significant reduction in the number states is necessary. The second issue is the choice of the precoding matrices and vectors. Recall that in this work we have assumed the choice of precoders to be pre-determined and fixed for each (state,action) pair. To fully exploit the precoding gains and the availability of ``precoded pilots" in future networks, we should relax this restriction. 
Finally, we remark that incorporating practical considerations such as delay constraints on scheduling are other  important open issues.


\newpage
 \appendices
\subsection{Extended MAT scheme}\label{sec:extEmat}

The \emph{MAT scheme} \cite{MAT} is an interesting tool that has been recently proposed to tackle the problem where no channel state estimates for the current interval are available at the BS but perfect albeit delayed CSI  is available to the BS. The scheme uses such completely outdated CSIT but still achieves  system degrees of freedom equal to $4/3$. We recall that in our context MU-MIMO with perfect and current CSIT will achieve 2 system degrees of freedom while single-user transmission will achieve only one degree of freedom.
In this paper, we will build upon the following extended MAT scheme \cite{Extended_MAT} that achieves the same system degrees of freedom as the original \emph{MAT scheme}. 

The scheme proceeds as follows. Time is divided into units referred to as rounds. 
Two messages ${\bm u}$ and ${\bm v}$ are to be transmitted, each destined to users $i$ and $j$ respectively, where ${\bm u}$ and ${\bm v}$ are $M_t\times 1$ vectors.
The three rounds are introduced next.

$\bullet$ \textbf{Round $1$}: The transmitted signal is $\bm x[1]={\bm u}+{\bm v}$, the corresponding received signal at user $i$ and $j$ is denoted by $y_i[1]$ and $y_j[1]$, where
{ {\begin{IEEEeqnarray}{rCl}
y_i[1]={{\bm h_{i}}}[1] ({\bm u}+{\bm v}) + n_i[1],\label{eq:Round1i} \\
y_j[1]={\bm h_{j}}[1] ({\bm u}+{\bm v}) +n_j[1].\label{eq:Round1j}
\end{IEEEeqnarray}}}
where $n_i[1]$ denotes the additive noise at user $i$ in round $1$ and ${{\bm h_{i}}}[1]\in\;\Cplx^{1\times M_t}$  denotes the channel response vector seen by user $i$ in Round $1$. 

$\bullet$ \textbf{Round $2$}: The transmitted signal is $\bm x[2]=[{{\bm h_{i}}}[1] {\bm v}; {\bm 0}]$, the received signal for user $i$ and $j$ is respectively
{ {\begin{IEEEeqnarray}{rCl}
y_i[2]=&h_{i,1}[2] \cdot ({{\bm h}_{i}}[1] {\bm v}) +n_i[2],\label{eq:Round21i} \\
y_j[2]=&h_{j,1}[2] \cdot ({{\bm h}_{i}}[1] {\bm v}) +n_j[2], \label{eq:Round21j}
\end{IEEEeqnarray}}}
where $h_{i,1}[2]$ denotes the channel coefficient modeling the propagation environment between user $i$ and the first transmit antenna at the BS during round $2$.

$\bullet$\textbf{Round $3$}: The transmitted signal is $\bm x_3=[{{\bm h_{j}}}[1] {\bm u}; {\bm 0}]$, the received signal for user $i$ and $j$ is respectively
{ {\begin{IEEEeqnarray}{rCl}
y_i[3]&=h_{i,1}[3] \cdot ({{\bm h}_{j}}[1]{\bm u}) +n_i[3], \label{eq:Round22i}\\
y_j[3]&=h_{j,1}[3] \cdot ({{\bm h}_{j}}[1] {\bm u}) +n_j[3]. \label{eq:Round22j}
\end{IEEEeqnarray}}}

It is assumed that the channel state vectors ${{\bm h_{i}}}[1]$, ${{\bm h_{i}}}[2]$, ${{\bm h_{i}}}[3]$ are estimated perfectly by user $i$ at the start of each respective round. Similarly for user $j$. In addition, the BS is assumed to know channel state vectors $\bm h_i[\ell],\bm h_j[\ell]$ perfectly but only after round $\ell$ for $\ell=1,2,3$. Further, user $i$ is also conveyed the channel vector ${\bm h}_{j}[1]$ and user $j$ is conveyed  the channel vector ${\bm h}_{i}[1]$ before the start of round 3, via feed-forward signaling. 

Therefore, after Round $3$, the $i^{th}$ user can decode message ${ \bm u}$ using (\ref{eq:Round1i}), (\ref{eq:Round21i})  and (\ref{eq:Round22i}) as per the following,
{ {\begin{IEEEeqnarray}{rCl}
y_i[1]-\frac{y_i[2]}{h_{i,1}[2]}&={{\bm h}_{i}}[1] {\bm u} + n_i[1]-\frac{n_i[2]}{h_{i,1}[2]}, \nonumber \\
y_i[3]&=h_{i,1}[3] \cdot {{\bm h}_{j}}[1] {\bm u} +n_i[3]. \nonumber
\end{IEEEeqnarray}}}
Similarly, after Round $3$, the $j^{th}$ user can also decode message ${\bm v}$. Notice that since the effective received observations seen by each user after three rounds can be modeled as the outputs of two linearly independent equations, each user can achieve two degrees of freedom over three rounds to attain system degrees of freedom equal to $4/3$.


\subsection{Proof of Proposition \ref{prop:P1}}\label{app:prop1}
{\begin{figure}\footnotesize{\begin{IEEEeqnarray}{rCl}
 Q_n^2[(\tau +1)T-1]\leq  \left(Q_n[\tau T]-\sum_{j=0}^{T-1}R^{\rm frame}_n[\tau T+j]\right)^2   + (Tr^*_n[\tau])^2 +
  2Tr^*_n[\tau]\left(Q_n[\tau T]-\sum_{j=0}^{T-1}R^{\rm frame}_n[\tau T+j]\right)^+\label{eq:bsq}
\end{IEEEeqnarray}}\end{figure}}
{\begin{figure}\footnotesize{\begin{IEEEeqnarray}{rCl}
 Q_n^2[(\tau +1)T] - (Q_n[\tau T])^2\leq\left(\sum_{j=0}^{T-1}R^{\rm frame}_n[\tau T+j]\right)^2   + (Tr^*_n[\tau])^2
  -2Q_n[\tau T]\left(\sum_{j=0}^{T-1}R^{\rm frame}_n[\tau T+j]-Tr^*_n[\tau]\right).\label{eq:bsq2}
\end{IEEEeqnarray}}\end{figure}}
  {\begin{figure}\scriptsize{\begin{IEEEeqnarray}{rCl}
\Delta_T({\bm Q}[\tau T])\leq \frac{1}{2T}E\left[\sum_{n=1}^N\left(\sum_{j=0}^{T-1}R^{\rm frame}_n[\tau T+j]\right)^2    + \sum_{n=1}^N(Tr^*_n[\tau])^2\middle\vert\bm Q[\tau T]\right]
 -\frac{1}{T}E\left[\sum_{n=1}^NQ_n[\tau T]\left(\sum_{j=0}^{T-1}R^{\rm frame}_n[\tau T+j]-Tr^*_n[\tau]\right)\middle\vert\bm Q[\tau T]\right]\label{eq:bsq3}\IEEEeqnarraynumspace
\end{IEEEeqnarray}}\end{figure}}
To bound the Lyapunov drift we proceed along the lines of \cite{KrishnaModiano} and first note that
{\begin{IEEEeqnarray}{rCl}
 Q_n[(\tau +1)T]\leq \left(Q_n[\tau T]-\sum_{j=0}^{T-1}R^{\rm frame}_n[\tau T+j]\right)^+  + Tr^*_n[\tau],\nonumber
\end{IEEEeqnarray}
so that (\ref{eq:bsq}) holds,
which then yields the bound in (\ref{eq:bsq2}).
Using (\ref{eq:bsq2}) we can bound  the $T-$step Lyapunov drift as in (\ref{eq:bsq3}).
 Then, since   $R^{\rm frame}_n[j],\forall\;n,j$ can be bounded above by a constant and $r^*_n[\tau]\leq r_{\rm max},\;\forall\;n,\tau$, we obtain the bound
 \begin{IEEEeqnarray}{rCl}
\Delta_T({\bm Q}[\tau T])\leq  BT + \sum_{n=1}^NQ_n[\tau T]r^*_n[\tau]\;\;\;\;\;\;\;\;\;\;\;\;\;\;\;\;\;\;\;\;\;\;\;\;\;\;\nonumber\\-E\left[\sum_{n=1}^NQ_n[\tau T]\frac{1}{T}\left(\sum_{j=0}^{T-1}R^{\rm frame}_n[\tau T+j]\right)
 \middle\vert\bm Q[\tau T]\right]\label{eq:bsq4}
\end{IEEEeqnarray}
where $B$ is an appropriate large enough constant.
The RHS in (\ref{eq:bsq4}) can be manipulated to obtain
 \begin{IEEEeqnarray}{rCl}
\Delta_T({\bm Q}[\tau T])\leq  BT + \sum_{n=1}^NQ_n[\tau T]r^*_n[\tau]-\sum_{n=1}^NQ_n[\tau T]R^*_n\;\;\;\;\;\;\;\;\;\;\;\;\;\;\nonumber\\
-E\left[\sum_{n=1}^NQ_n[\tau T]\left(\frac{1}{T}\left(\sum_{j=0}^{T-1}R^{\rm frame}_n[\tau T+j]\right)- R^*_n\right)
 \middle\vert\bm Q[\tau T]\right]\nonumber
\end{IEEEeqnarray}
where $\bm R^*=[R^*_1,\cdots,R_N^*]^T$ was defined in (\ref{eq:Rnstar}). Using the Cauchy-Schwartz inequality along with the fact that $\sum_{n=1}^NQ_n[\tau T]\geq \sqrt{\sum_{n=1}^NQ_n^2[\tau T]}$, we can then further upper bound
 \begin{IEEEeqnarray}{rCl}
\Delta_T({\bm Q}[\tau T])\leq  BT + \sum_{n=1}^NQ_n[\tau T]r^*_n[\tau]-\sum_{n=1}^NQ_n[\tau T]R^*_n+\;\;\;\;\;\;\;\;\;\;\;\;\;\;\;\nonumber\\
(\sum_{n=1}^NQ_n[\tau T])E\left[\left\|\frac{1}{T}\left(\sum_{j=0}^{T-1}\bm R^{\rm frame}[\tau T+j]\right)- \bm R^*\right\| \middle\vert\bm Q[\tau T]\right]\label{eq:bsq5}\IEEEeqnarraynumspace
\end{IEEEeqnarray}
Invoking Lemma \ref{lem:mixing} along with the fact that ${\bm R}^*$ is also bounded above, we can deduce that by choosing a large enough frame length we can ensure that
\begin{IEEEeqnarray}{rCl}
 E\left[\left\|\frac{1}{T}\left(\sum_{j=0}^{T-1}\bm R^{\rm frame}[\tau T+j]\right)- \bm R^*\right\| \middle\vert\bm Q[\tau T]\right]\leq \epsilon
\end{IEEEeqnarray}
which when used in (\ref{eq:bsq5}) yields
\begin{IEEEeqnarray}{rCl}
  \Delta_T({\bm Q}[\tau T])\leq  BT + \sum_{n=1}^NQ_n[\tau T]r^*_n[\tau]-\sum_{n=1}^NQ_n[\tau T]R^*_n+\nonumber\\
\epsilon\sum_{n=1}^NQ_n[\tau T].\;\;\;\;\;\;
\end{IEEEeqnarray}
 Recall that any vector $\bm R$ in the $\epsilon-$interior of $\Lambda$ satisfies $\bm R\preceq\tilde{\bm R}-\epsilon{\bm 1}$ for some $\tilde{\bm R}\in\Lambda$. Then, appealing to the fact that $\sum_{n=1}^NQ_n[\tau T]R^*_n$ is the optimal solution for the LP in (\ref{eq:LP}) together with Lemma \ref{prop:region}, we have that
 \begin{IEEEeqnarray}{rCl}
  \Delta_T({\bm Q}[\tau T])\leq  BT + \sum_{n=1}^NQ_n[\tau T]r^*_n[\tau]-\sum_{n=1}^NQ_n[\tau T](\tilde{R}_n
-\epsilon),\nonumber
\end{IEEEeqnarray}
 from which (\ref{eq:Tub}) follows.

 \subsection{Proof Sketch of Theorem \ref{prop:P2}}\label{app:thm1}
 We leverage some of the techniques used in \cite {Caire_fb3} but we emphasize that the policies considered in \cite{Caire_fb3} were not frame based and Markov decision processes were not employed there.
Using the result in (\ref{eq:Tub}) (after assuming a large enough framelength) and subtracting the term $VU(\bm r^*[\tau])$ from both sides, we first obtain
 { {\begin{IEEEeqnarray}{rCl}
\Delta_T({\bm Q}[\tau T])-VU(\bm r^*[\tau])\leq BT  - \sum_{n=1}^N Q_n[\tau T]R_n + \sum_{n=1}^N Q_n[\tau T]r^*_n[\tau]-VU(\bm r^*[\tau]).
\end{IEEEeqnarray}}}
Then recalling that $\bm r^*[\tau]$ is the optimal solution to (\ref{eq:Aopt}) we have that for any
$\bm v: \bm 0\preceq \bm v\preceq r_{\rm max}{\bm 1}$
{ {\begin{IEEEeqnarray}{rCl}
\Delta_T({\bm Q}[\tau T])-VU(\bm r^*[\tau])\leq BT  - \sum_{n=1}^N Q_n[\tau T]R_n +\nonumber\\  \sum_{n=1}^N Q_n[\tau T]v_n-VU(\bm v).\label{eq:pp2}
\end{IEEEeqnarray}}}
Averaging both sides of (\ref{eq:pp2}) with respect to $\bm Q[\tau T]$, we obtain
{ {\begin{IEEEeqnarray}{rCl}
\frac{1}{T}E[L(\bm Q[(\tau+1) T])]-\frac{1}{T}E[L(\bm Q[\tau T])] -V E[U(\bm r^*[\tau])]\leq BT\;\;\;\;\;\;\;\;\;\;\;\;\;\;\;\;\;\;\;\;\;\;\;\;\;\;\;\;\;\;\;\;\;\; \nonumber\\ - \sum_{n=1}^N E[Q_n[\tau T]]R_n +  \sum_{n=1}^N E[Q_n[\tau T]]v_n-VU(\bm v).\IEEEeqnarraynumspace\label{eq:pp3}
\end{IEEEeqnarray}}}
Noting that $Q_n[0]=0,\;\forall\; n$ and summing (\ref{eq:pp3}) over $\tau=0,1,\cdots, t-1$ we get
{ {\begin{IEEEeqnarray}{rCl}
\frac{1}{T}E[L(\bm Q[t T])] - \sum_{\tau=0}^{t-1}VE[U(\bm r^*[\tau])]\leq BTt \;\;\;\;\;\;\;\;\;\;\;\;\;\;\;\;\;\;\;\;\;\;\;\;\;\;\;\;\;\;\;\;\;\; \;\;\;\;\;\;\;\;\;\;\;\;\;\;\;\;\;\;\;\;\;\;\;\;\;\;\;\nonumber\\  - \sum_{n=1}^N \sum_{\tau=0}^{t-1}E[Q_n[\tau T]]R_n+  \sum_{n=1}^N \sum_{\tau=0}^{t-1}E[Q_n[\tau T]]v_n-tVU(\bm v)\nonumber
\end{IEEEeqnarray}}}
which when combined with the fact that $\frac{1}{T}E[L(\bm Q[t T])]\geq 0$ yields
{ {\begin{IEEEeqnarray}{rCl}
\frac{1}{t}\sum_{n=1}^N \sum_{\tau=0}^{t-1}E[Q_n[\tau T]](R_n-v_n)\leq BT + \frac{1}{t} \sum_{\tau=0}^{t-1}VE[U(\bm r^*[\tau])]\nonumber\\ -VU(\bm v).\IEEEeqnarraynumspace\label{eq:pp4}
\end{IEEEeqnarray}}}
Next, choosing any $\bm R \in\Lambda_{\epsilon}$ and $\bm v: \bm 0\prec \bm v=\bm R-\delta\bm 1$ and $\bm v\preceq r_{\rm max} \bm 1$ for some $\delta>0$, and substituting in (\ref{eq:pp4}), we get that
{ {\begin{IEEEeqnarray}{rCl}
\frac{1}{t}\sum_{n=1}^N \sum_{\tau=0}^{t-1}\delta E[Q_n[\tau T]]\leq BT + \frac{1}{t} \sum_{\tau=0}^{t-1}VE[U(\bm r^*[\tau])] -VU(\bm v),\nonumber
\end{IEEEeqnarray}}}
which using the componentwise non-increasing property of the utility function yields
{ {\begin{IEEEeqnarray}{rCl}
\frac{1}{t}\sum_{n=1}^N \sum_{\tau=0}^{t-1}\delta E[Q_n[\tau T]]\leq BT + VU(r_{\rm max}\bm 1) -VU(\bm v),\;\forall\;t.\IEEEeqnarraynumspace\label{eq:pp3ft}
\end{IEEEeqnarray}}}
Then, since $Q_n[\tau T+j]\leq Q_n[\tau T] + jr_{\rm max},\;\forall\;n,j$ and $U(\bm v)\geq\vartheta>-\infty$ for some constant $\vartheta$, from (\ref{eq:pp3ft}) we can conclude that
$\frac{1}{J}\sum_{n=1}^N \sum_{j=0}^{J-1}\delta E[Q_n[j]]$ is also bounded above by a constant for all $J$,
which proves that all virtual queues are strongly stable under the frame based policy. Letting $A_n[\tau T+ j]={\bm r}^*[\tau],\;\forall\;0\leq j\leq T-1,\tau=0,1,\cdots$, denote the per-slot virtual arrival rate, from strong stability of each virtual queue, uniformly bounded arrival rates and uniform continuity of the utility function, we can deduce that
{ {\begin{IEEEeqnarray}{rCl}
\lim\inf_{J\to\infty}U\left(\frac{1}{J} \sum_{j=0}^{J-1}E\left[\bm A[j]\right]\right)\leq \;\;\;\;\;\;\;\;\;\;\;\;\;\;\;\;\;\;\;\;\nonumber\\ \lim\inf_{J\to\infty}U\left(\frac{1}{J} \sum_{j=0}^{J-1}E\left[\bm R^{\rm frame}[j]\right]\right).\IEEEeqnarraynumspace\label{eq:limineq}
\end{IEEEeqnarray}}}

Finally, setting $\bm R=\bm v=\bm r_{\epsilon}^{\rm opt}$ in (\ref{eq:pp4}), we obtain
{ {\begin{IEEEeqnarray}{rCl}
 \frac{1}{t} \sum_{\tau=0}^{t-1}VE[U(\bm r^*[\tau])] \geq VU(\bm r_{\epsilon}^{\rm opt}) -BT,
\end{IEEEeqnarray}}}
which upon invoking the concavity of the utility function and the linearity of the expectation operator yields
{ {\begin{IEEEeqnarray}{rCl}
 U\left(\frac{1}{t} \sum_{\tau=0}^{t-1}E\left[\bm r^*[\tau]\right] \right)\geq U(\bm r_{\epsilon}^{\rm opt}) -BT/V.\label{eq:firstLB}
\end{IEEEeqnarray}}}

Notice then that due to the uniform continuity of the utility function, $\lim\inf_{t\to \infty}U\left(\frac{1}{t} \sum_{\tau=0}^{t-1}E\left[\bm r^*[\tau]\right]\right)$ is equal to
{ {\begin{IEEEeqnarray}{rCl}
 \lim\inf_{J\to \infty}U\left(\frac{1}{TJ} \sum_{j=0}^{TJ-1}E\left[\bm A[j]\right]\right)\;\;\;\;\;\;\;\;\;\;\;\;\nonumber\\
 = \lim\inf_{J\to \infty}U\left(\frac{1}{J} \sum_{j=0}^{J-1}E\left[\bm A[j]\right]\right)  \label{eq:pplimI}\IEEEeqnarraynumspace
\end{IEEEeqnarray}}}
which when used in (\ref{eq:firstLB}) yields
{ {\begin{IEEEeqnarray}{rCl}
 \lim\inf_{J\to \infty} U\left(\frac{1}{J} \sum_{j=0}^{J-1}E\left[\bm A[j]\right]\right)   \geq U(\bm r_{\epsilon}^{\rm opt}) -BT/V.\label{eq:pplim}\IEEEeqnarraynumspace
\end{IEEEeqnarray}}}
Using (\ref{eq:pplim})  and (\ref{eq:limineq}) yields the desired result.\endproof

 \newpage

\begin{figure}
\centering
\includegraphics[width=4in]{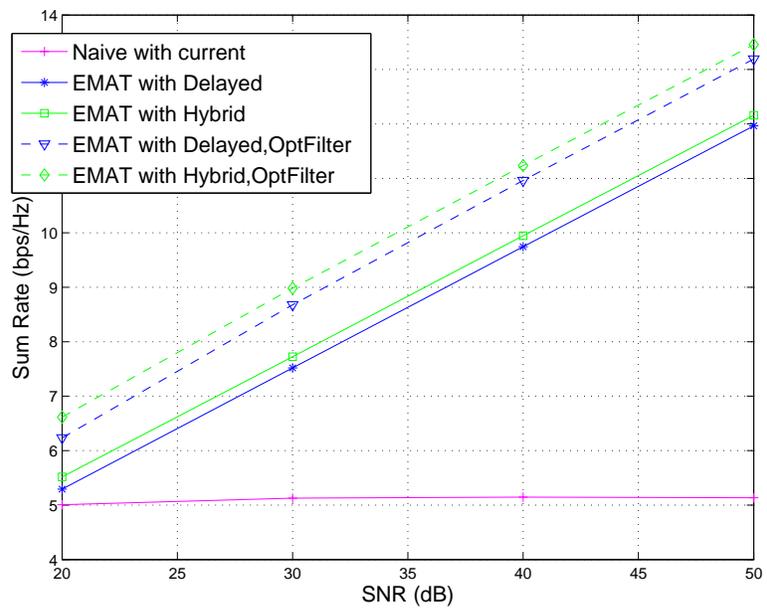}
\caption{Comparison with conventional MU-MIMO}
\label{fig:simul}
\end{figure}

\newpage
\newpage
\begin{figure}\vspace{5cm}
\centering
\includegraphics[width=4in]{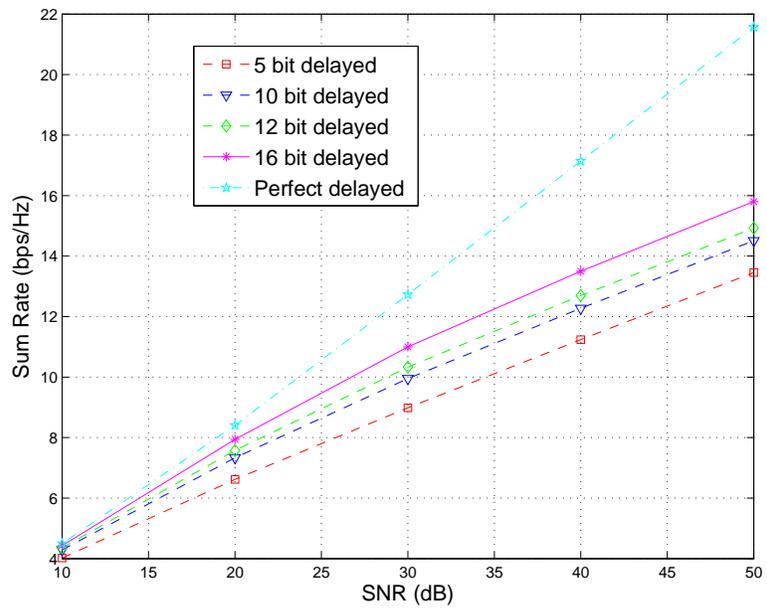}
\caption{Comparison for different codebook sizes}
\label{fig:simul2}
\end{figure}

\end{document}